\documentstyle[11pt]{article}
\begin{document}
\title{Self-Duality of Various Chiral Boson Actions}
\author{{Yan-Gang Miao${}^{{\rm a,b},1,2}$ and H.J.W. M$\ddot{\rm u}$ller-Kirsten
${}^{{\rm a},3}$}\\
{\small ${}^{\rm a}$ Department of Physics, University of Kaiserslautern,
P.O. Box 3049,}\\
{\small D-67653 Kaiserslautern, Germany}\\
{\small ${}^{\rm b}$ Department of Physics, Xiamen University, Xiamen 361005,}\\
{\small People's Republic of China}}
\date{}
\maketitle
\footnotetext[1]{Alexander von Humboldt Fellow.} 
\footnotetext[2]{E-mail: miao@physik.uni-kl.de}
\footnotetext[3]{E-mail: mueller1@physik.uni-kl.de}
\vskip 48pt
\begin{center}{\bf Abstract}\end{center}
\baselineskip 22pt
    The duality symmetries of various chiral boson actions are investigated     using
$D=2$ and $D=6$ space-time dimensions as examples. These actions involve the
Siegel, Floreanini-Jackiw, Srivastava and Pasti-Sorokin-Tonin formulations. We
discover that the Siegel, Floreanini-Jackiw and Pasti-Sorokin-Tonin actions
have
self-duality with respect to a common anti-dualization of chiral boson fields
in $D=2$ and $D=6$ dimensions, respectively, while the Srivastava action is
self-dual with respect to a generalized dualization of chiral boson fields.
Moreover, the action of the Floreanini-Jackiw chiral bosons interacting with
gauge fields in $D=2$ dimensions also has self-duality but with respect to a
generalized anti-dualization of chiral boson fields.
\vskip 24pt
PACS number(s): 11.10.Kk
\newpage
\section{Introduction}
\par
Chiral {\em p}-forms, sometimes called chiral bosons, are described by an
antisymmetric {\em p}th order tensor $A^{(p)}$ in the $D=2(p+1)$ dimensional
space-time, whose
external differential $F^{(p+1)}(A)=dA^{(p)}$ satisfies the self-duality
condition
\begin{equation}
{\cal F}^{(p+1)} \equiv F^{(p+1)}(A)-{}^{\ast}F^{(p+1)}(A)=0,
\end{equation}
where ${}^{\ast}F^{(p+1)}(A)$ is
defined as the dual
partner of $F^{(p+1)}(A)$. In the
space with the
Lorentzian metric signature, the
self-duality requires $A^{(p)}$ to be
real if {\em p} is
even, or complex if {\em p} is odd. In the latter case the theory can 
equivalently be described by a pair of real antisymmetric tensor fields related
by a duality condition.

    Chiral bosons have attracted much attention because they play an important
role in many theoretical models. In $D=2$ dimensional space-time, they occur as
basic ingredients and elements in the formulation of heterotic strings [1] and
in a number of statistical systems [2]. In $D>2$ dimensional space-time, they
form an integral part in $D=6$ and type
IIB $D=10$ supergravity and M-theory
five-branes [3-6]. Since the equation of motion of a chiral boson, i.e., the
self-duality condition, is first order with respect to the derivatives of
space and time, it is a key problem to construct the corresponding action and
then to quantize the theory consistently. To this end, various formulations of
actions have been proposed [7-12]. These actions can be classified by
manifestly Lorentz covariant versions [7-10] and non-manifestly Lorentz
covariant versions [11,12] when one emphasizes their formalism under the
Lorentz transformation, or by polynomial versions [7-9] and non-polynomial
version [10] when one focuses on auxiliary fields introduced in the actions.
Incidentally, there are no auxiliary fields introduced in the non-manifestly
Lorentz covariant actions [11,12].

    Many proposals have been suggested to construct chiral boson actions,
among which are four typical ones [7,11,8,10] we are interested in here. The
first scheme, proposed by Siegel [7], is to impose the square of the
self-duality condition upon a {\em p}th
order
antisymmetric tensor field through the
introduction of an auxiliary tensor field as a Lagrange multiplier. The
problem is that the Siegel action suffers from an anomaly of gauge
symmetries. However, it is possible [7] to cancel the anomaly either by
introducing a Liouville term or by taking a system of 26 chiral bosons. The
second proposal, by Floreanini and Jackiw [11] only in $D=2$ dimensions, is to
offer a unitary and Poincar${\acute{\rm e}}$
invariant
formulation by means of a first order
Lagrangian in the following three ways: (i) a nonlocal Lagrangian in terms of
a local field, (ii) a local Lagrangian in terms of a nonlocal field, and (iii)
a local Lagrangian in terms of a local field which is of fermionic
character. The equivalence between item (ii),
known as the Floreanini-Jackiw
formulation, and the Siegel formulation in $D=2$ dimensions has been shown by
Bernstein and Sonnenschein [13], and the intrinsic relation between items (i)
and (iii) has also been uncovered by
Girotti {\em et al.}[14] from the point
of view
of chiral bosonization. In addition, the Floreanini-Jackiw formulation has
been generalized to $D=2(p+1)$ dimensional
space-time by Henneaux {\em et al.}[12].
The third proposal, suggested by
Srivastava [8] by following Siegel's idea but
adding the self-duality condition itself, gives rise to the so-called linear
formulation of chiral bosons in $D=2$ dimensions. Although it has some defects
as pointed out by Harada [15] and
Girotti {\em et al.}[16], the linear
formulation
strictly describes a chiral boson from the point of view of equations of
motion at both the classical and quantum levels. Moreover, it is quite
straightforward to generalize this
formulation to $D=2(p+1)$ dimensional
space-time (cf. Subsect.4.2). The fourth
scheme, recently proposed by Pasti,
Sorokin and Tonin [10], is to construct a Lorentz covariant formulation of
chiral {\em p}-forms in $D=2(p+1)$ dimensions that contains a finite number of
auxiliary fields in a non-polynomial way. The simplest case is that only one
auxiliary scalar field is introduced. This formulation reduces to the
non-manifestly covariant Floreanini-Jackiw formulation [11] provided
appropriate gauge fixing conditions are chosen. On the other hand, it has a
close relationship with the Lorentz covariant McClain-Wu-Yu formulation [9]
that contains infinitely many auxiliary fields in the usual polynomial way.
That is to say, the Pasti-Sorokin-Tonin formulation turns into the
McClain-Wu-Yu formulation if one gets rid of the non-polynomiality and
eliminates the scalar auxiliary field at the price of introducing auxiliary
$(p+1)$-forms, or, vice versa, if one consistently truncates the McClain-Wu-Yu
infinite tail and puts on its end the auxiliary scalar field.

    Because various types of strings are related by dualities, the duality
symmetries of the Pasti-Sorokin-Tonin formulation have been studied and
some interesting results have been
obtained [10]. The chiral boson action in
$D=2$ dimensions is self-dual with respect to both the dualization of the chiral
boson field and the dualization of the auxiliary scalar field. In the $D=4$
case, the action is still self-dual under the dualization of the two real
chiral 1-forms, but turns out to be a new covariant duality-symmetric Maxwell
action that contains an auxiliary 2-form field under a duality transform of
the auxiliary scalar field. The Pasti-Sorokin-Tonin action in $D=6$ dimensional
space-time gives rise to such a dual version that includes an auxiliary 4-form
field and has a different symmetry structure from that of its initial action
when one performs a duality transform of the auxiliary scalar field.
Incidentally, the self-duality of the action with respect to the dualization
of the chiral 2-form field in the $D=6$ case was not explicitly verified in
Ref.[10].

    In this paper we investigate the duality properties of the four typical
chiral {\em p}-form actions mentioned above by using $D=2$ and $D=6$ dimensions as
examples. We pay our main attention to
these actions' dual versions under
duality transforms of chiral {\em p}-form fields since we expect to extract some
common property from the four actions that have such big differences in
formulation. As to the duality under transforms of auxiliary fields for the
first three chiral {\em p}-form actions, it is a trivial problem because of the
linearity of auxiliary fields in the Siegel and Srivastava actions [7,8] and
of the non-existence of auxiliary fields in the Floreanini-Jackiw action
[11,12]. As a result, we discover that the Siegel, Floreanini-Jackiw and
Pasti-Sorokin-Tonin actions are self-dual under a common anti-dual transform
of 1-form `field strengths' in $D=2$ dimensional space-time and of 3-form field
strengths in the $D=6$ case, while the Srivastava action is self-dual under a
generalized dual transform of 1-form
`field strength' in $D=2$ dimensions and
of 3-form field strength in $D=6$ dimensions. We also find that the
self-duality conditions of the four actions in the $D=2$ and $D=6$ cases,
respectively, have the same transformation although the transforms of the
field strengths are quite different from one another. Moreover, we extend the
self-duality of actions from free chiral bosons to interacting cases and
choose, as an example, the action of the Floreanini-Jackiw chiral bosons
interacting with gauge fields proposed by Harada [17]. We find that this
action is also self-dual but with respect to a generalized anti-dualization of
the chiral boson field, and that the transformation of the difference between
the 1-form `field strength' and its dual partner is very different from that of
the free cases because of interactions.

    The paper is arranged as follows. In Sects. 2, 3,
4 and 5, we discuss the duality symmetries of the four chiral {\em p}-form actions
one by one in the Siegel, Floreanini-Jackiw, Srivastava and
Pasti-Sorokin-Tonin formulations. Each section is divided into two subsections
for the $D=2$ and $D=6$ cases. Then we turn to the interacting theory of the
Floreanini-Jackiw chiral bosons and gauge fields in Sect. 6, and finally make
a conclusion in Sect.7.

    The metric notation we use throughout this paper is
\begin{eqnarray}
g_{00}=-g_{11}=\cdots=-g_{D-1,D-1}=1,\nonumber \\
{\epsilon}^{012{\cdots}D-1}=1.
\end{eqnarray}
Greek letters stand for space-time
indices (${{\mu},{\nu},{\sigma},\cdots
=0,1,\cdots,D-1)}$
and Latin letters are spacial indices running from 1 to $D-1$.
\section{Self-duality of the Siegel action}
\subsection{The D=2 case}
\par
    We begin with the Siegel action [7] in $D=2$ dimensional space-time
\begin{equation}
S=\int d^{2}x\left[{\frac 1 2}{\partial}_{\mu}{\phi}{\partial}^{\mu}{\phi}
+{\frac 1 2}{\lambda}_{{\mu}{\nu}}\left({\partial}^{\mu}{\phi}-
{\epsilon}^{{\mu}{\sigma}}{\partial}_{\sigma}{\phi}\right)
\left({\partial}^{\nu}{\phi}-
{\epsilon}^{{\nu}{\rho}}{\partial}_{\rho}{\phi}\right)\right],
\end{equation}
where $\phi$ is a scalar field, and ${\lambda}_{{\mu}{\nu}}$ a symmetric and
traceless auxiliary
tensor field.

   We investigate the duality property of eq.(3) with respect to the
dualization of the field ${\phi}(x)$
along the line of Ref.[10]. The first step is to
introduce two independent vector fields, $F_{\mu}$ and $G_{\mu}$, and
replace eq.(3) by the action
\begin{equation}
S=\int d^{2}x\left[{\frac 1 2}F_{\mu}F^{\mu}+{\frac 1 2}{\lambda}_{{\mu}
{\nu}}{\cal F}^{\mu}{\cal F}^{\nu}+G^{\mu}\left(F_{\mu}-{\partial}_{\mu}
{\phi}\right)\right] ,
\end{equation}
where ${\cal F}^{\mu}$ is defined as the difference between $F^{\mu}$ and its
dual partner ${\epsilon}^{{\mu}{\nu}}F_{\nu}$
\begin{equation}
{\cal F}^{\mu}=F^{\mu}-{\epsilon}^{{\mu}{\nu}}F_{\nu}.
\end{equation}
Then, varying eq.(4) with respect to $G^{\mu}$ gives the expression for the
field $F_{\mu}$ in terms of $\phi$
\begin{equation}
F_{\mu}={\partial}_{\mu}{\phi},
\end{equation}
together with which eq.(4) turns back to the original Siegel action eq.(3).
This shows the classical equivalence between actions eqs.(3) and (4). The third
step is to vary eq.(4) with respect to $F_{\mu}$, which yields the expression
of $G^{\mu}$ in terms of $F^{\mu}$
\begin{equation}
G^{\mu}=-F^{\mu}-\left(g^{{\mu}{\sigma}}+{\epsilon}^{{\mu}{\sigma}}\right)
{\lambda}_{{\sigma}{\nu}}{\cal F}^{\nu}.
\end{equation}
Similar to eq.(5), ${\cal G}^{\mu}$ is defined as
\begin{equation}
{\cal G}^{\mu}=G^{\mu}-{\epsilon}^{{\mu}{\nu}}G_{\nu},
\end{equation}
which, when eq.(7) is substituted, gives a relationship between
${\cal F}^{\mu}$ and ${\cal G}^{\mu}$
\begin{equation}
{\cal F}^{\mu}=-{\cal G}^{\mu}.
\end{equation}
It is necessary to point out in advance that eq.(9) is generally satisfied for
all the four chiral boson actions discussed in this paper (cf. Subsects. 3.1,
4.1 and 5.1) although the relations between $F^{\mu}$ and $G^{\mu}$ for these
actions
are very different from one another. With eq.(9), it is easy to invert eq.(7)
and obtain $F^{\mu}$ in terms of $G^{\mu}$
\begin{equation}
F^{\mu}=-G^{\mu}+\left(g^{{\mu}{\sigma}}+{\epsilon}^{{\mu}{\sigma}}\right)
{\lambda}_{{\sigma}{\nu}}{\cal G}^{\nu}.
\end{equation}
We can check from eq.(7) that when the self-duality condition is satisfied,
i.e., ${\cal F}^{\mu}=0$, which is called `on the mass shell' in Ref.[10],
$F^{\mu}$ and $G^{\mu}$ relate with an anti-duality $G^{\mu}=-{\epsilon}
^{{\mu}{\nu}}F_{\nu}$.
Note that in Ref.[10] they
relate with a dual relation because of the distinct metric notation. We will
see that this type of anti-duality also appears in the Floreanini-Jackiw and
Pasti-Sorokin-Tonin actions in the D=2 case although eqs.(7), (27) and (51)
are
quite different from one another (cf. Subsects. 3.1 and 5.1), but does not in
the Srivastava action (cf. Subsect. 4.1). Substituting eq.(10) into eq.(4), we
get the dual version of the Siegel action
\begin{equation}
S_{dual}=\int d^{2}x\left[-{\frac 1 2}G_{\mu}G^{\mu}+{\frac 1 2}
{\lambda}_{{\mu}
{\nu}}{\cal G}^{\mu}{\cal G}^{\nu}+{\phi}{\partial}_{\mu}
G^{\mu}\right].
\end{equation}
Variation of eq.(11) with respect to $\phi$ gives
${\partial}_{\mu}G^{\mu}=0$,
whose solution should be
\begin{equation}
G^{\mu}(\psi)=-{\epsilon}^{{\mu}{\nu}}
{\partial}_{\nu}{\psi}\equiv
-{\epsilon}^{{\mu}{\nu}}F_{\nu}(\psi),
\end{equation}
where $\psi$ is an arbitrary scalar
field.
When eq.(12) is substituted
into eq.(11), we obtain the dual action that is exactly the same as the Siegel
action eq.(3) only with the replacement
of $\phi$ by $\psi$. As analysed above,
$\phi$
and $\psi$ coincide with each other up
to a constant when the self-duality
condition is imposed. Therefore, the Siegel action is self-dual with respect
to the ${\phi}(x) - {\psi}(x)$
anti-dualization expressed by eqs.(6)
and (12).
\subsection{The D=6 case}
\par
    The Siegel action in $D=6$ space-time dimensions takes the form [7]
\begin{equation}
S=\int d^{6}x\left[{\frac 1 6}F_{{\mu}{\nu}{\sigma}}(A)
F^{{\mu}{\nu}{\sigma}}(A)+{\frac 1 2}{\lambda}_{{\mu}{\nu}}
{\cal F}^{{\mu}{\rho}{\sigma}}(A){{\cal F}^{\nu}}_{{\rho}{\sigma}}(A)\right],
\end{equation}
where $F_{{\mu}{\nu}{\sigma}}(A)$ is the 3-form field strength of
the real antisymmetric tensor field $A_{{\mu}{\nu}} (\mu,\nu =0,1,\cdots,5)$
\begin{equation}
F_{{\mu}{\nu}{\sigma}}(A)={\partial}_{\mu}A_{{\nu}{\sigma}}+
{\partial}_{\nu}A_{{\sigma}{\mu}}+{\partial}_{\sigma}A_{{\mu}{\nu}}
\equiv {\partial}_{[{\mu}}A_{{\nu}{\sigma}]},
\end{equation}
and ${\cal F}_{{\mu}{\nu}{\sigma}}(A)$ is defined as
\begin{equation}
{\cal F}_{{\mu}{\nu}{\sigma}}(A)=F_{{\mu}{\nu}{\sigma}}(A)
-{\frac{1}{3!}}{\epsilon}_{{\mu}{\nu}{\sigma}{\rho}{\eta}{\delta}}
F^{{\rho}{\eta}{\delta}}(A).
\end{equation}
\par
    In order to discuss the duality of the Siegel action, we introduce two
3-form fields $F_{{\mu}{\nu}{\sigma}}$ and $G_{{\mu}{\nu}{\sigma}}$,
and replace eq.(13) by the following action
\begin{equation}
S=\int d^{6}x\left[{\frac 1 6}F_{{\mu}{\nu}{\sigma}}
F^{{\mu}{\nu}{\sigma}}+{\frac 1 2}{\lambda}_{{\mu}{\nu}}
{\cal F}^{{\mu}{\rho}{\sigma}}{{\cal F}^{\nu}}_{{\rho}{\sigma}}
+{\frac 1 3}G^{{\mu}{\nu}{\sigma}}\left(F_{{\mu}{\nu}{\sigma}}
-{\partial}_{[{\mu}}A_{{\nu}{\sigma}]}\right)\right],
\end{equation}
where $F_{{\mu}{\nu}{\sigma}}$ and $G_{{\mu}{\nu}{\sigma}}$ act, at present, as
independent auxiliary fields. To vary eq.(16) with respect to
$G^{{\mu}{\nu}{\sigma}}$ gives
\begin{equation}
F_{{\mu}{\nu}{\sigma}}={\partial}_{[{\mu}}A_{{\nu}{\sigma}]},
\end{equation}
which, when substituted into eq.(16), yields the equivalence between actions
eqs.(13) and (16). On the other hand, variation of eq.(16) with respect to
$F_{{\mu}{\nu}{\sigma}}$ leads to the expression of $G^{{\mu}{\nu}{\sigma}}$
in terms of $F^{{\mu}{\nu}{\sigma}}$
\begin{equation}
G^{{\mu}{\nu}{\sigma}}=-F^{{\mu}{\nu}{\sigma}}
-{\lambda}^{{\rho}[{\mu}}{{\cal F}_{\rho}}^{{\nu}{\sigma}]}
-{\frac{1}{3!}}{\epsilon}^{{\mu}{\nu}{\sigma}{\rho}{\eta}{\delta}}
{\lambda}_{{\theta}[{\rho}}{{\cal F}^{\theta}}_{{\eta}{\delta}]}.
\end{equation}
Like eq.(15), we define ${\cal G}^{{\mu}{\nu}{\sigma}}$ to be
\begin{equation}
{\cal G}^{{\mu}{\nu}{\sigma}}=G^{{\mu}{\nu}{\sigma}}
-{\frac{1}{3!}}{\epsilon}^{{\mu}{\nu}{\sigma}{\rho}{\eta}{\delta}}
G_{{\rho}{\eta}{\delta}},
\end{equation}
and obtain, when eq.(18) is substituted into eq.(19), the relation
\begin{equation}
{\cal F}^{{\mu}{\nu}{\sigma}}=-{\cal G}^{{\mu}{\nu}{\sigma}}.
\end{equation}
Note that this is generally satisfied for all the four actions in the $D=6$ case
although relations of $F^{{\mu}{\nu}{\sigma}}$ and $G^{{\mu}{\nu}{\sigma}}$ in
these actions are quite different
from one another (cf. Subsects. 3.2, 4.2, and 5.2). With eq.(20), we can invert
eq.(18) quite easily and solve $F^{{\mu}{\nu}{\sigma}}$ in terms
of $G^{{\mu}{\nu}{\sigma}}$
\begin{equation}
F^{{\mu}{\nu}{\sigma}}=-G^{{\mu}{\nu}{\sigma}}
+{\lambda}^{{\rho}[{\mu}}{{\cal G}_{\rho}}^{{\nu}{\sigma}]}
+{\frac{1}{3!}}{\epsilon}^{{\mu}{\nu}{\sigma}{\rho}{\eta}{\delta}}
{\lambda}_{{\theta}[{\rho}}{{\cal G}^{\theta}}_{{\eta}{\delta}]}.
\end{equation}
We can verify from eq.(18) that when the self-duality condition is satisfied,
i.e., ${\cal F}^{{\mu}{\nu}{\sigma}}=0$, $F^{{\mu}{\nu}{\sigma}}$
and $G^{{\mu}{\nu}{\sigma}}$
relate with an anti-duality
$G^{{\mu}{\nu}{\sigma}}=
-{\frac{1}{3!}}{\epsilon}^{{\mu}{\nu}{\sigma}{\rho}{\eta}{\delta}}
F_{{\rho}{\eta}{\delta}}$.
This relation also appears
in the Floreanini-Jackiw and Pasti-Sorokin-Tonin actions in the $D=6$ case, but
does not in the Srivastava action. Now substituting eq.(21) into the action
eq.(16), we obtain the dual Siegel action in the $D=6$ case
\begin{equation}
S_{dual}=\int d^{6}x\left[-{\frac 1 6}G_{{\mu}{\nu}{\sigma}}
G^{{\mu}{\nu}{\sigma}}+{\frac 1 2}{\lambda}_{{\mu}{\nu}}
{\cal G}^{{\mu}{\rho}{\sigma}}{{\cal G}^{\nu}}_{{\rho}{\sigma}}
+A_{{\nu}{\sigma}}{\partial}_{\mu}G^{{\mu}{\nu}{\sigma}}\right].
\end{equation}
Variation of eq.(22) with respect to $A_{{\nu}{\sigma}}$
gives
\begin{equation}
{\partial}_{\mu}G^{{\mu}{\nu}{\sigma}}=0,
\end{equation}
whose solution should be
\begin{equation}
G^{{\mu}{\nu}{\sigma}}(B)=
-{\frac{1}{3!}}{\epsilon}^{{\mu}{\nu}{\sigma}{\rho}{\eta}{\delta}}
{\partial}_{[{\rho}}B_{{\eta}{\delta}]}
\equiv
-{\frac{1}{3!}}{\epsilon}^{{\mu}{\nu}{\sigma}{\rho}{\eta}{\delta}}
F_{{\rho}{\eta}{\delta}}(B),
\end{equation}
where $B_{{\mu}{\nu}}$ is an arbitrary
2-form field. When eq.(24) is
substituted into the dual action eq.(22), we get the result that the dual
action is the same as the Siegel action eq.(13) only with the replacement of
$A_{{\mu}{\nu}}$ by $B_{{\mu}{\nu}}$.
Consequently, the Siegel
action is self-dual in $D=6$ dimensional
space-time with respect to the
$A_{{\mu}{\nu}} - B_{{\mu}{\nu}}$
anti-dualization given
by eqs.(17) and (24).
\section{Self-duality of the Floreanini-Jackiw action}
\subsection{The D=2 case}
\par
    The Floreanini-Jackiw action in $D=2$ dimensions has the form [11]
\begin{equation}
S=\int
d^{2}x\left[{\partial}_{0}{\phi}{\partial}_{1}{\phi}-({\partial}_{1}{\phi})^2
\right],
\end{equation}
in which no auxiliary fields are introduced. It is a non-manifestly Lorentz
covariant action, but has Poincar$\acute{\rm e}$ invariance from the point of
view of Hamiltonian analyses.

    As in Subsect. 2.1, we introduce two independent auxiliary vector
fields $F^{\mu}$ and $G^{\mu}$, and replace eq.(25) by the action
\begin{equation}
S=\int
d^{2}x\left[F_{0}F_{1}-(F_{1})^2+G^{\mu}(F_{\mu}-{\partial}_{\mu}{\phi})
\right].
\end{equation}
Variation of this action with respect to the Lagrange multiplier
$G^{\mu}$ gives
rise to the same result as eq.(6), which leads to the equivalence between eqs.(25)
and (26). On the other hand, variation of eq.(26) with respect to $F_{\mu}$
gives the expression of $G^{\mu}$ in terms of $F_{\mu}$
\begin{eqnarray}
G^{0}&=&-F_{1},\nonumber \\
G^{1}&=&-F_{0}+2F_{1},
\end{eqnarray}
whose inversion is
\begin{eqnarray}
F_{0}&=&-2G^{0}-G^{1},\nonumber \\
F_{1}&=&-G^{0}.
\end{eqnarray}
If we define ${\cal F}^{\mu}$ and ${\cal G}^{\mu}$ as in eqs.(5) and (8),
respectively,
we discover that they still satisfy the relation eq.(9) as pointed out in
Subsect. 2.1. Moreover, $F^{\mu}$ and $G^{\mu}$ have an anti-dual relation
$G^{\mu}=-{\epsilon}^{{\mu}{\nu}}F_{\nu}$ if the self-duality condition in the
$D=2$ case ${\cal F}^{\mu}=0$
is imposed into eq.(27). Substituting eq.(28) into eq.(26), we
obtain the dual Floreanini-Jackiw action
\begin{equation}
S_{dual}=\int d^{2}x\left[-(G_{0})^2+G_{0}G_{1}+{\phi}{\partial}_{\mu}G^{\mu}
\right].
\end{equation}
The remaining procedure is the same as that in Subsect. 2.1. As a result, the
Floreanini-Jackiw action in $D=2$ dimensional space-time is self-dual with
respect to the ${\phi}(x) - {\psi}(x)$ anti-duality as shown in
eqs.(6) and (12).
\subsection{The D=6 case}
\par
    The non-manifestly Lorentz covariant formulation of Floreanini and Jackiw
was generalized to chiral {\em p}-forms in Ref.[12]. The action for a chiral 2-form
in $D=6$ dimensions is
\begin{equation}
S=\int d^{6}x\left[{\frac 1 2}\left(F_{0ij}(A)-{\frac{1}{3!}}{\epsilon}
_{0ijklm}F^{klm}(A)\right)\cdot{\frac{1}{3!}}{\epsilon}_{0ijnpq}F^{npq}(A)
\right],
\end{equation}
where $F_{{\mu}{\nu}{\sigma}}(A)$ is the field strength of $A_{{\mu}{\nu}}$, as
stated
in eq.(14), and Latin letters stand for spatial indices
($i,j,\cdots=1,\cdots,5$). Note
that no auxiliary fields appear in eq.(30). In the following, we utilize the
simplier form of eq.(30)
\begin{equation}
S=\int d^{6}x\left[{\frac{1}{12}}{\epsilon}^{0ijklm}F_{0ij}(A)F_{klm}(A)
-{\frac 1 6}F_{klm}(A)F_{klm}(A)\right].
\end{equation}
\par
We begin with the duality property of the action eq.(31) under the
dualization of the antisymmetric tensor field $A_{{\mu}{\nu}}$.
Introducing two auxiliary 3-forms $F_{{\mu}{\nu}{\sigma}}$ and
$G_{{\mu}{\nu}{\sigma}}$,
we construct a new action to replace eq.(31)
\begin{equation}
S=\int d^{6}x\left[{\frac{1}{12}}{\epsilon}^{0ijklm}F_{0ij}F_{klm}
-{\frac 1 6}F_{klm}F_{klm}+{\frac 1 6}G^{{\mu}{\nu}{\sigma}}
(F_{{\mu}{\nu}{\sigma}}-{\partial}_{[{\mu}}A_{{\nu}{\sigma}]})\right],
\end{equation}
where $F_{{\mu}{\nu}{\sigma}}$ and $G_{{\mu}{\nu}{\sigma}}$ are treated as
independent fields. For the sake of
convenience in the calculation, we rewrite eq.(32) to be
\begin{eqnarray}
S=\int d^{6}x\left[{\frac{1}{12}}{\epsilon}^{0ijklm}F_{0ij}F_{klm}
-{\frac 1 6}F_{klm}F_{klm}
+{\frac 1 2}G^{0ij}(F_{0ij}-{\partial}_{[0}A_{ij]})\right. \nonumber \\
\left.+{\frac 1 6}G^{klm}
(F_{klm}-{\partial}_{[k}A_{lm]})\right].
\end{eqnarray}
Variation of eq.(33) with respect to the
Lagrange multiplier $G^{{\mu}{\nu}{\sigma}}$ gives
eq.(17), which shows the equivalence between eqs.(31) and (33). Moreover, variation
of eq.(33) with respect to $F_{{\mu}{\nu}{\sigma}}$ gives the
expression of $G^{{\mu}{\nu}{\sigma}}$
in terms of $F_{{\mu}{\nu}{\sigma}}$
\begin{eqnarray}
G^{0ij}&=&-{\frac 1 6}{\epsilon}^{0ijnpq}F_{npq},\nonumber \\
G^{klm}&=&-{\frac 1 2}{\epsilon}^{0npklm}F_{0np}-2F^{klm},
\end{eqnarray}
from which $F_{{\mu}{\nu}{\sigma}}$ can be calculated
\begin{eqnarray}
F^{0ij}&=&-2G^{0ij}+{\frac 1 6}{\epsilon}^{0ijnpq}G_{npq},\nonumber \\
F^{klm}&=&{\frac 1 2}{\epsilon}^{0npklm}G_{0np}.
\end{eqnarray}
If we define ${\cal F}^{{\mu}{\nu}{\sigma}}$ and
${\cal G}^{{\mu}{\nu}{\sigma}}$ as in eqs.(15) and (19), respectively, we
find that they still satisfy eq.(20) although eqs.(18) and (34), i.e., the
relations of $F^{{\mu}{\nu}{\sigma}}$ and $G^{{\mu}{\nu}{\sigma}}$ for the
Siegel and Floreanini-Jackiw
formulations of chiral 2-forms, are quite different. In addition, when
imposing the self-duality condition in the $D=6$ case, i.e.,
${\cal F}^{{\mu}{\nu}{\sigma}}=0$,
into eq.(34), we still derive the anti-duality between
$F^{{\mu}{\nu}{\sigma}}$ and $G^{{\mu}{\nu}{\sigma}}$,
$G^{{\mu}{\nu}{\sigma}}=
-{\frac{1}{3!}}{\epsilon}^{{\mu}{\nu}{\sigma}{\rho}{\eta}{\delta}}
F_{{\rho}{\eta}{\delta}}$.
Substituting
eq.(35) into eq.(33), we obtain the dual formulation of the Floreanini-Jackiw
chiral 2-form in the $D=6$ case
\begin{equation}
S_{dual}=\int d^{6}x\left[-{\frac 1 2}G^{0ij}G_{0ij}
+{\frac{1}{12}}{\epsilon}_{0ijklm}G^{0ij}G^{klm}
+{\frac 1 2}A_{{\nu}{\sigma}}{\partial}_{\mu}
G^{{\mu}{\nu}{\sigma}}\right].
\end{equation}
The following steps are straightforward. Variation of eq.(36) with
respect to $A_{{\nu}{\sigma}}$ gives
${\partial}_{\mu}G^{{\mu}{\nu}{\sigma}}=0$,
whose solution is eq.(24) in
which an antisymmetric tensor field $B_{{\mu}{\nu}}$ is introduced. With
eq.(24), the
dual action eq.(36) is the same as the action eq.(31), only with the replacement of
$A_{{\mu}{\nu}}$ by $B_{{\mu}{\nu}}$. Therefore, we verify
that the Floreanini-Jackiw action
for a chiral 2-form in $D=6$ dimensions is self-dual under the $A_{{\mu}{\nu}}
- B_{{\mu}{\nu}}$ anti-duality transform of eqs.(17) and (24).
\section{Self-duality of the Srivastava action}
\subsection{The D=2 case}
\par
    We write the linear formulation of chiral bosons suggested by Srivastava
[8]
\begin{equation}
S=\int d^{2}x\left[{\frac 1 2}{\partial}_{\mu}{\phi}{\partial}^{\mu}{\phi}
+{\lambda}_{\mu}({\partial}^{\mu}{\phi}-{\epsilon}^{{\mu}{\nu}}
{\partial}_{\nu}{\phi})\right],
\end{equation}
where $\phi$ is a scalar field and ${\lambda}_{\mu}$ an auxiliary vector
field. This
action has some defects as pointed out by others [15,16], but it `synthesizes' the manifest
Lorentz covariance and self-duality constraint.

    Let us introduce two auxiliary vector fields $F_{\mu}$ and $G^{\mu}$,
and construct a new action to replace eq.(37)
\begin{equation}
S=\int d^{2}x\left[{\frac 1 2}F^{\mu}F_{\mu}+{\lambda}_{\mu}(F^{\mu}
-{\epsilon}^{{\mu}{\nu}}F_{\nu})+G^{\mu}(F_{\mu}-{\partial}_{\mu}{\phi})
\right],
\end{equation}
where $F_{\mu}$ and $G^{\mu}$ are independent of the other fields. When varying
eq.(38)
with respect to $G^{\mu}$, we have $F_{\mu}={\partial}_{\mu}{\phi}$, i.e.,
eq.(6), and with
it we can prove that the new action eq.(38) is equivalent to the original one
eq.(37). On the other hand, when varying eq.(38) with respect to $F_{\mu}$,
we get $G^{\mu}$ in terms of $F_{\mu}$
\begin{equation}
G^{\mu}=-F^{\mu}-({\lambda}^{\mu}+{\epsilon}^{{\mu}{\nu}}{\lambda}_{\nu}),
\end{equation}
or, vice versa, $F_{\mu}$ in terms of $G^{\mu}$
\begin{equation}
F^{\mu}=-G^{\mu}-({\lambda}^{\mu}+{\epsilon}^{{\mu}{\nu}}{\lambda}_{\nu}).
\end{equation}
If ${\cal F}^{\mu}$ and ${\cal G}^{\mu}$ are defined as in eqs.(5) and (8),
respectively, they again
satisfy the relation eq.(9) although eq.(39) is quite different from eq.(7)
and eq.(27). However, $F^{\mu}$ and $G^{\mu}$ no longer relate with any
anti-duality when the self-duality condition ${\cal F}^{\mu}=0$ is imposed
into eq.(39). This happens because the self-duality condition with a Lagrange
multiplier is introduced linearly in the action eq.(37). We may say that this
anti-duality between $F^{\mu}$ and $G^{\mu}$ is not necessary when one
considers the
duality property of actions because the self-duality condition can not be
directly imposed into actions. Substituting eq.(40) into eq.(38), we obtain
the dual version of the Srivastava action
\begin{equation}
S_{dual}=\int d^{2}x\left[-{\frac 1 2}G^{\mu}G_{\mu}-{\lambda}_{\mu}(G^{\mu}
-{\epsilon}^{{\mu}{\nu}}G_{\nu})+{\phi}{\partial}_{\mu}G^{\mu}
\right].
\end{equation}
When varying eq.(41) with respect to $\phi$, we get
${\partial}_{\mu}G^{\mu}=0$ and then solve
\begin{equation}
G^{\mu}(\psi)={\epsilon}^{{\mu}{\nu}}{\partial}_{\nu}{\psi}
\equiv {\epsilon}^{{\mu}{\nu}}F_{\nu}(\psi),
\end{equation}
where $\psi$ is an arbitrary scalar field. When eq.(42) is
substituted into eq.(41), we find that the dual action is the same as the       original
one eq.(37) only with the replacement of $\phi$ by $\psi$. Consequently, the
Srivastava action in the $D=2$ case is self-dual under the generalized
dualization eq.(42). Here the word `generalized' means that ${\phi}(x)$ does
not coincide with ${\psi}(x)$ even if the self-duality condition is
considered.
\subsection{The D=6 case}
\par
    We can easily generalize the $D=2$ Srivastava action to the $D=6$ case
\begin{equation}
S=\int d^{6}x\left[{\frac 1 6}F_{{\mu}{\nu}{\sigma}}(A)
F^{{\mu}{\nu}{\sigma}}(A)+{\frac 1 3}{\lambda}_{{\mu}{\nu}{\sigma}}
{\cal F}^{{\mu}{\nu}{\sigma}}(A)\right],
\end{equation}
where $F_{{\mu}{\nu}{\sigma}}(A)$ and ${\cal F}^{{\mu}{\nu}{\sigma}}(A)$ are
defined as eqs.(14) and (15),
respectively, and ${\lambda}_{{\mu}{\nu}{\sigma}}$ is an auxiliary
antisymmetric tensor field.
Variation of this action with respect to ${\lambda}_{{\mu}{\nu}{\sigma}}$
gives the self-duality
condition ${\cal F}^{{\mu}{\nu}{\sigma}}(A)=0$ that is in fact the equation of
motion
of $A_{{\mu}{\nu}}$. Therefore, eq.(43) indeed describes a chiral 2-form field
in $D=6$
dimensional space-time. As to its canonical Hamiltonian analysis, it can be
achieved straightforwardly by following the procedure shown in Ref.[8]. Here
we omit it.

    We introduce two auxiliary 3-form fields $F_{{\mu}{\nu}{\sigma}}$ and
$G_{{\mu}{\nu}{\sigma}}$, and construct a new action to replace eq.(43)
\begin{equation}
S=\int d^{6}x\left[{\frac 1 6}F_{{\mu}{\nu}{\sigma}}
F^{{\mu}{\nu}{\sigma}}+{\frac 1 3}{\lambda}_{{\mu}{\nu}{\sigma}}
{\cal F}^{{\mu}{\nu}{\sigma}}+{\frac 1 3}G^{{\mu}{\nu}{\sigma}}
(F_{{\mu}{\nu}{\sigma}}-{\partial}_{[{\mu}}A_{{\nu}{\sigma}]})\right],
\end{equation}
where $F_{{\mu}{\nu}{\sigma}}$ and $G^{{\mu}{\nu}{\sigma}}$ are treated as
independent fields, and
${\cal F}^{{\mu}{\nu}{\sigma}}\equiv F^{{\mu}{\nu}{\sigma}}
-\frac{1}{3!}{\epsilon}^{{\mu}{\nu}{\sigma}{\rho}{\eta}{\delta}}
F_{{\rho}{\eta}{\delta}}$.
By
varying eq.(44) with respect to $G^{{\mu}{\nu}{\sigma}}$, we get
$F_{{\mu}{\nu}{\sigma}}={\partial}_{[{\mu}}A_{{\nu}{\sigma}]}$ and
then verify the equivalence between eqs.(43) and (44). On the other hand, by
varying eq.(44) with respect to $F_{{\mu}{\nu}{\sigma}}$, we have the
expression of $G^{{\mu}{\nu}{\sigma}}$ in terms of $F_{{\mu}{\nu}{\sigma}}$
\begin{equation}
G^{{\mu}{\nu}{\sigma}}=-F^{{\mu}{\nu}{\sigma}}-({\lambda}^{{\mu}{\nu}{\sigma}}
+\frac{1}{3!}{\epsilon}^{{\mu}{\nu}{\sigma}{\rho}{\eta}{\delta}}
{\lambda}_{{\rho}{\eta}{\delta}}),
\end{equation}
or, vice versa, that of $F_{{\mu}{\nu}{\sigma}}$ in terms of
$G_{{\mu}{\nu}{\sigma}}$
\begin{equation}
F^{{\mu}{\nu}{\sigma}}=-G^{{\mu}{\nu}{\sigma}}-({\lambda}^{{\mu}{\nu}{\sigma}}
+\frac{1}{3!}{\epsilon}^{{\mu}{\nu}{\sigma}{\rho}{\eta}{\delta}}
{\lambda}_{{\rho}{\eta}{\delta}}).
\end{equation}
As usual, we define
${\cal G}^{{\mu}{\nu}{\sigma}}=G^{{\mu}{\nu}{\sigma}}
-\frac{1}{3!}{\epsilon}^{{\mu}{\nu}{\sigma}{\rho}{\eta}{\delta}}
G_{{\rho}{\eta}{\delta}}$
and obtain, using eq.(45),
${\cal F}^{{\mu}{\nu}{\sigma}}=-{\cal G}^{{\mu}{\nu}{\sigma}}$.
This relation is generally correct for all the four formulations of chiral
2-forms although eqs.(18), (34), (45) and (56) are quite different from one
another. But, similar to the $D=2$ case, $F^{{\mu}{\nu}{\sigma}}$ and
$G^{{\mu}{\nu}{\sigma}}$ do not
relate with any anti-duality in the Srivastava action even if the self-duality
condition ${\cal F}^{{\mu}{\nu}{\sigma}}=0$ is imposed to eq.(45). The reason
remains the
linearity of the self-duality condition in the action eq.(43). This situation
does not occur in the Siegel, Floreanini-Jackiw and Pasti-Sorokin-Tonin
actions. Substituting eq.(46) into eq.(44), we get the dual action
\begin{equation}
S_{dual}=\int d^{6}x\left[-{\frac 1 6}G_{{\mu}{\nu}{\sigma}}
G^{{\mu}{\nu}{\sigma}}-{\frac 1 3}{\lambda}_{{\mu}{\nu}{\sigma}}
{\cal G}^{{\mu}{\nu}{\sigma}}+A_{{\nu}{\sigma}}{\partial}_{{\mu}}
G^{{\mu}{\nu}{\sigma}}\right].
\end{equation}
Variation of eq.(47) with respect to $A_{{\nu}{\sigma}}$ gives
${\partial}_{{\mu}}G^{{\mu}{\nu}{\sigma}}=0$, and the solution should be
\begin{equation}
G^{{\mu}{\nu}{\sigma}}(B)=
{\frac{1}{3!}}{\epsilon}^{{\mu}{\nu}{\sigma}{\rho}{\eta}{\delta}}
{\partial}_{[{\rho}}B_{{\eta}{\delta}]}
\equiv
{\frac{1}{3!}}{\epsilon}^{{\mu}{\nu}{\sigma}{\rho}{\eta}{\delta}}
F_{{\rho}{\eta}{\delta}}(B),
\end{equation}
where $B_{{\mu}{\nu}}$ is an arbitrary 2-form field. Substituting
eq.(48) into eq.(47), we recover the Srivastava formulation with
$B_{{\mu}{\nu}}$ as the
argument. This shows the self-duality of the Srivastava action in the $D=6$ case
with respect to the generalized duality transform eq.(48). Here we add the
word `generalized' because $A_{{\mu}{\nu}}$ no longer coincides with
$B_{{\mu}{\nu}}$
on the mass shell.
\section{Self-duality of the Pasti-Sorokin-Tonin action}
\subsection{The D=2 case}
\par
    The self-duality of the Pasti-Sorokin-Tonin action in the $D=2$ case has
been explicitly shown in Ref.[10]. In order to make our paper complete, we
briefly repeat the main procedure by means of our metric notation that is
different from that used in Ref.[10].

    The non-polynomial formulation of chiral bosons proposed by Pasti, Sorokin
and Tonin [10] takes the form
\begin{equation}
S=\int d^{2}x\left\{{\frac 1 2}{\partial}_{\mu}{\phi}{\partial}^{\mu}{\phi}
+\frac{1}{2({\partial}_{\nu}a)({\partial}^{\nu}a)}\left[{\partial}^{\mu}a
({\partial}_{\mu}{\phi}-{\epsilon}_{{\mu}{\sigma}}{\partial}^{\sigma}{\phi})
\right]^2\right\},
\end{equation}
where ${\phi}(x)$ is a scalar field, and $a(x)$ an auxiliary scalar field
introduced in a non-polynomial way. Note that we have adopted our metric
notation in the action eq.(49).

    By introducing two auxiliary vector fields $F_{\mu}$ and $G^{\mu}$, we
construct a new action to replace eq.(49)
\begin{equation}
S=\int d^{2}x\left[{\frac 1 2}F_{\mu}F^{\mu}
+\frac{1}{2({\partial}_{\nu}a)({\partial}^{\nu}a)}
({\partial}^{\mu}a{\cal F}_{\mu})^2+G^{\mu}(F_{\mu}-{\partial}_{\mu}{\phi})
\right],
\end{equation}
where $F_{\mu}$ and $G^{\mu}$ are dealt with as independent fields, and
${\cal F}_{\mu}\equiv F_{\mu}-{\epsilon}_{{\mu}{\nu}}F^{\nu}$.
Variation of eq.(50) with respect to the Lagrange multiplier $G^{\mu}$ gives
$F_{\mu}={\partial}_{\mu}{\phi}$, i.e., eq.(6), which yields the equivalence
between the
Pasti-Sorokin-Tonin action and the new action eq.(50). Moreover, variation of
eq.(50) with respect to $F_{\mu}$ leads to the expression of $G^{\mu}$ in
terms of $F_{\mu}$
\begin{equation}
G^{\mu}=-F^{\mu}-\frac{{\partial}^{\mu}a+{\epsilon}^{{\mu}{\rho}}{\partial}
_{\rho}a}{({\partial}_{\sigma}a)({\partial}^{\sigma}a)}({\partial}^{\nu}a
{\cal F}_{\nu}).
\end {equation}
In order to easily solve $F_{\mu}$ in terms of $G^{\mu}$ from the above
equation, we define, like eq.(8), ${\cal G}^{\mu}=G^{\mu}-{\epsilon}
^{{\mu}{\nu}}G_{\nu}$.
When eq.(51) is substituted into ${\cal G}^{\mu}$, we get the relation ${\cal F}^{\mu}=
-{\cal G}^{\mu}$,
which also exists in the first three formulations of chiral bosons discussed
in Subsects 2.1, 3.1 and 4.1. By using ${\cal F}^{\mu}=-{\cal G}^{\mu}$,
we therefore solve $F^{\mu}$ from eq.(51)
\begin{equation}
F^{\mu}=-G^{\mu}+\frac{{\partial}^{\mu}a+{\epsilon}^{{\mu}{\rho}}{\partial}
_{\rho}a}{({\partial}_{\sigma}a)({\partial}^{\sigma}a)}({\partial}^{\nu}a
{\cal G}_{\nu}).
\end {equation}
We can see that $F^{\mu}$ and $G^{\mu}$ satisfy an anti-duality $G^{\mu}=-
{\epsilon}^{{\mu}{\nu}}F_{\nu}$
on the mass shell. Note that in Ref.[10] their relation is dual because of the
distinct metric notation. We have known that this type of anti-duality also
appears in the Siegel and Floreanini-Jackiw actions in the $D=2$ case although
eqs.(7), (27) and (51) are quite different from one another, but does not in
the Srivastava action. Now substituting eq.(52) into eq.(50), we obtain the
dual action
\begin{equation}
S_{dual}=\int d^{2}x\left[-{\frac 1 2}G_{\mu}G^{\mu}
+\frac{1}{2({\partial}_{\nu}a)({\partial}^{\nu}a)}
({\partial}^{\mu}a{\cal G}_{\mu})^2+{\phi}{\partial}_{\mu}G^{\mu}
\right].
\end{equation}
Exactly following the discussions below eq.(11), we can conclude that the
Pasti-Sorokin-Tonin action in $D=2$ dimensional space-time is self-dual with
respect to the ${\phi}(x) - {\psi}(x)$ anti-dualization given by eqs.(6) and
(12).
\subsection{The D=6 case}
\par
    Since the self-duality of the Pasti-Sorokin-Tonin action with respect to
the dualization of chiral 2-form fields in $D=6$ dimensional space-time was not
explicitly verified in Ref.[10], we add the verification here in terms of our
metric notation.

    First we write the Pasti-Sorokin-Tonin action for a chiral 2-form
field $A_{{\mu}{\nu}}$
\begin{equation}
S=\int d^{6}x\left[{\frac 1 6}F_{{\mu}{\nu}{\sigma}}(A)F^{{\mu}{\nu}{\sigma}}
(A)
+\frac{1}{2({\partial}_{\lambda}a)({\partial}^{\lambda}a)}
{\partial}^{\mu}a{\cal F}_{{\mu}{\nu}{\sigma}}(A)
{\cal F}^{{\nu}{\sigma}{\rho}}(A){\partial}_{\rho}a
\right],
\end{equation}
where $F_{{\mu}{\nu}{\sigma}}(A)$ and ${\cal F}_{{\mu}{\nu}{\sigma}}(A)$ are
defined as in eqs.(14) and (15),
respectively, and $a(x)$ is an auxiliary scalar field introduced in a
non-polynomial way.

    By introducing two auxiliary 3-form fields $F_{{\mu}{\nu}{\sigma}}$ and
$G_{{\mu}{\nu}{\sigma}}$, we construct a new action to replace eq.(54)
\begin{eqnarray}
S=\int d^{6}x\left[{\frac 1 6}F_{{\mu}{\nu}{\sigma}}F^{{\mu}{\nu}{\sigma}}
+\frac{1}{2({\partial}_{\lambda}a)({\partial}^{\lambda}a)}
{\partial}^{\mu}a{\cal F}_{{\mu}{\nu}{\sigma}}
{\cal F}^{{\nu}{\sigma}{\rho}}{\partial}_{\rho}a\right.\nonumber \\
\left.+{\frac 1 3}
G^{{\mu}{\nu}{\sigma}}(F_{{\mu}{\nu}{\sigma}}-{\partial}_{[{\mu}}
A_{{\nu}{\sigma}]})
\right],
\end{eqnarray}
where $F_{{\mu}{\nu}{\sigma}}$ and $G_{{\mu}{\nu}{\sigma}}$ are dealt with as
independent fields, and
${\cal F}^{{\mu}{\nu}{\sigma}}\equiv F^{{\mu}{\nu}{\sigma}}
-\frac{1}{3!}{\epsilon}^{{\mu}{\nu}{\sigma}{\rho}{\eta}{\delta}}
F_{{\rho}{\eta}{\delta}}$.
Variation of eq.(55) with respect to the Lagrange multiplier
$G^{{\mu}{\nu}{\sigma}}$
gives $F_{{\mu}{\nu}{\sigma}}={\partial}_{[{\mu}}A_{{\nu}{\sigma}]}$, i.e.,
eq.(17), which yields the
equivalence between eqs.(54) and (55). On the other hand, variation of eq.(55) with
respect to $F_{{\mu}{\nu}{\sigma}}$ leads to the expression of
$G_{{\mu}{\nu}{\sigma}}$ in terms of $F_{{\mu}{\nu}{\sigma}}$
\begin{equation}
G^{{\mu}{\nu}{\sigma}}=-F^{{\mu}{\nu}{\sigma}}
-\frac{1}{({\partial}_{\lambda}a)({\partial}^{\lambda}a)}
\left[{\partial}^{[{\mu}}a{\cal F}^{{\nu}{\sigma}]{\rho}}
{\partial}_{\rho}a
+\frac{1}{3!}{\epsilon}^{{\mu}{\nu}{\sigma}{\rho}{\eta}{\delta}}
{\partial}_{[{\rho}}a{\cal F}_{{\eta}{\delta}]{\theta}}
{\partial}^{\theta}a\right].
\end{equation}
When we define
${\cal G}^{{\mu}{\nu}{\sigma}}=G^{{\mu}{\nu}{\sigma}}
-\frac{1}{3!}{\epsilon}^{{\mu}{\nu}{\sigma}{\rho}{\eta}{\delta}}
G_{{\rho}{\eta}{\delta}}$,
we obtain ${\cal F}^{{\mu}{\nu}{\sigma}}=-{\cal G}^{{\mu}{\nu}{\sigma}}$ once
again. As
we have pointed out in Subsect. 5.1, this relation is generally correct for all the
four chiral 2-form actions in $D=6$ dimensions although eqs.(18), (34), (45) and
(56) are quite different from one another. Considering the general relation,
we can solve from eq.(56) $F^{{\mu}{\nu}{\sigma}}$ in terms
of $G^{{\mu}{\nu}{\sigma}}$
\begin{equation}
F^{{\mu}{\nu}{\sigma}}=-G^{{\mu}{\nu}{\sigma}}
+\frac{1}{({\partial}_{\lambda}a)({\partial}^{\lambda}a)}
\left[{\partial}^{[{\mu}}a{\cal G}^{{\nu}{\sigma}]{\rho}}
{\partial}_{\rho}a
+\frac{1}{3!}{\epsilon}^{{\mu}{\nu}{\sigma}{\rho}{\eta}{\delta}}
{\partial}_{[{\rho}}a{\cal G}_{{\eta}{\delta}]{\theta}}
{\partial}^{\theta}a\right].
\end{equation}
As discussed in Subsects. 2.2 and 3.2, we can prove that
$G^{{\mu}{\nu}{\sigma}}$ relates to $F^{{\mu}{\nu}{\sigma}}$
by an anti-duality
$G^{{\mu}{\nu}{\sigma}}=
-\frac{1}{3!}{\epsilon}^{{\mu}{\nu}{\sigma}{\rho}{\eta}{\delta}}
F_{{\rho}{\eta}{\delta}}$
on the mass shell, that is, under the condition
${\cal F}^{{\mu}{\nu}{\sigma}}=0$.
The anti-dual relation is satisfied in the Siegel, Floreanini-Jackiw and
Pasti-Sorokin-Tonin actions, but not in the Srivastava action. Now
substituting eq.(57) into eq.(55), we get the dual action in terms of
$G^{{\mu}{\nu}{\sigma}}$
\begin{equation}
S_{dual}=\int d^{6}x\left[-{\frac 1 6}
G_{{\mu}{\nu}{\sigma}}G^{{\mu}{\nu}{\sigma}}
+\frac{1}{2({\partial}_{\lambda}a)({\partial}^{\lambda}a)}
{\partial}^{\mu}a{\cal G}_{{\mu}{\nu}{\sigma}}
{\cal G}^{{\nu}{\sigma}{\rho}}{\partial}_{\rho}a
+A_{{\nu}{\sigma}}{\partial}_{{\mu}}
G^{{\mu}{\nu}{\sigma}}
\right].
\end{equation}
We do not repeat the subsequent steps which are equally the same as  below
eq.(22). As a result, the Pasti-Sorokin-Tonin action has self-duality under
the $A_{{\mu}{\nu}} - B_{{\mu}{\nu}}$ anti-dual transform
eqs.(17) and (24).
\section{Self-duality of the gauged Floreanini-Jackiw chiral boson action}
\par
    We extend the discussion of self-duality of chiral {\em p}-form actions from
free theories to interacting cases, and choose the action of Floreanini-Jackiw
chiral bosons interacting with gauge fields [17] as our example.

    We first write the action of this interacting theory
\begin{eqnarray}
S=\int d^{2}x\left[{\partial}_{0}{\phi}{\partial}_{1}{\phi}
-({\partial}_{1}{\phi})^2+2e{\partial}_{1}{\phi}(A_{0}-A_{1})\right.\nonumber
\\
\left.-{\frac 1 2}e^2(A_{0}-A_{1})^2+{\frac 1 2}
e^2aA_{\mu}A^{\mu} -{\frac 1 4}F_{{\mu}{\nu}}F^{{\mu}{\nu}}\right],
\end{eqnarray}
where $\phi$ is a scalar field, $A_{\mu}$ a gauge field and $F_{{\mu}{\nu}}$
its field strength;
{\em e} is the electric charge and {\em a} a real parameter caused by
ambiguity in
bosonization. It is a non-manifestly Lorentz covariant action but indeed has
Lorentz invariance [17]. In the following discussion, the interacting term,
i.e., the third term in eq.(59), is important, while the last three terms that
relate only to gauge fields have nothing to do with the duality property
of the action.

    By introducing two auxiliary vector fields $F_{\mu}$ and $G^{\mu}$, we
construct a new action to replace eq.(59)
\begin{eqnarray}
S=\int d^{2}x\left[F_{0}F_{1}
-(F_{1})^2+2eF_{1}(A_{0}-A_{1})
-{\frac 1 2}e^2(A_{0}-A_{1})^2\right.\nonumber \\
\left.+{\frac 1 2}e^2aA_{\mu}A^{\mu}
-{\frac 1 4}F_{{\mu}{\nu}}F^{{\mu}{\nu}}
+G^{\mu}(F_{\mu}-{\partial}_{\mu}{\phi})\right],
\end{eqnarray}
where $F_{\mu}$ and $G^{\mu}$ are treated as independent fields. Variation of
eq.(60) with respect to the Lagrange multiplier $G^{\mu}$ gives
$F_{\mu}={\partial}_{\mu}{\phi}$,
which yields the equivalence between the two actions eqs.(59) and (60).
Furthermore, variation of eq.(60) with respect to $F_{\mu}$ leads to the
expression of $G^{\mu}$ in terms of $F_{\mu}$
\begin{eqnarray}
G^{0}&=&-F_{1},\nonumber \\
G^{1}&=&-F_{0}+2F_{1}-2e(A_{0}-A_{1}).
\end{eqnarray}
It is easy to solve for $F_{\mu}$ from the above equation
\begin{eqnarray}
F_{0}&=&-2G^{0}-G^{1}-2e(A_{0}-A_{1}),\nonumber \\
F_{1}&=&-G^{0}.
\end{eqnarray}
If we define ${\cal F}_{\mu}=F_{\mu}-{\epsilon}_{{\mu}{\nu}}F^{\nu}$ and
${\cal G}_{\mu}=G_{\mu}-{\epsilon}_{{\mu}{\nu}}G^{\nu}$,
we find that they satisfy the relation
\begin{equation}
{\cal F}_{\mu}=-{\cal G}_{\mu}-2e(g_{{\mu}{\nu}}-{\epsilon}_{{\mu}{\nu}})
A^{\nu},
\end{equation}
which is different from that of the free Floreanini-Jackiw case because of
interactions. In other words, if the interaction did not exist, i.e., $e=0$,
eq.(63) would reduce to the free theory case
${\cal F}_{\mu}=-{\cal G}_{\mu}$.
Substituting eq.(62) into eq.(60), we obtain the dual action in terms of
$G^{\mu}$
\begin{eqnarray}
S_{dual}=\int d^{2}x\left[-(G^{0})^2
-G^{0}G^{1}-2eG^{0}(A_{0}-A_{1})
-{\frac 1 2}e^2(A_{0}-A_{1})^2\right.\nonumber \\
\left.+{\frac 1 2}e^2aA_{\mu}A^{\mu}
-{\frac 1 4}F_{{\mu}{\nu}}F^{{\mu}{\nu}}
+{\phi}{\partial}_{\mu}G^{\mu}\right].
\end{eqnarray}
Variation of eq.(64) with respect to $\phi$ gives ${\partial}_{\mu}G^{\mu}=0$,
whose solution should be
\begin{equation}
G^{\mu}(\psi)=-{\epsilon}^{{\mu}{\nu}}{\partial}_{\nu}{\psi}
\equiv -{\epsilon}^{{\mu}{\nu}}F_{\nu}({\psi}),
\end{equation}
where ${\psi}(x)$ is an arbitrary scalar field. Substituting eq.(65) into
eq.(64), we get the dual action in terms of $\psi$
\begin{eqnarray}
S_{dual}=\int d^{2}x\left[{\partial}_{0}{\psi}{\partial}_{1}{\psi}
-({\partial}_{1}{\psi})^2+2e{\partial}_{1}{\psi}(A_{0}-A_{1})\right.\nonumber
\\
\left.-{\frac 1 2}e^2(A_{0}-A_{1})^2+{\frac 1 2}e^2aA_{\mu}A^{\mu}
-{\frac 1 4}F_{{\mu}{\nu}}F^{{\mu}{\nu}}\right].
\end{eqnarray}
It has the same formulation as the original action eq.(59) only with the
replacement of $\phi$ by $\psi$. Note that because of interactions, ${\phi}(x)$
no longer coincides with ${\psi}(x)$ up to a constant on the mass shell, which
is
different from that of the free theory case. This means that eq.(65) shows a
generalized anti-dualization of $F_{\mu}$ and $G_{\mu}$. Therefore, we prove
that the action of gauged Floreanini-Jackiw chiral bosons has self-duality
with respect to the generalized anti-dualization of `field strength'
expressed by eq.(65). Incidentally, if we chose the solution
$G^{\mu}(\psi)={\epsilon}^{{\mu}{\nu}}{\partial}_{\nu}{\psi}$
instead of eq.(65), the dual action would have a minus sign in the third term.
That is to say, the dual action derived in this way would be different from
the action eq.(59) in formulation. However, the physical spectrum is the same
whether the third term of eq.(66) is positive or negative.
\section{Conclusion}
\par
    By following the procedure of duality analyses illustrated by Pasti,
Sorokin and Tonin [10], we have proved that the Siegel, Floreanini-Jackiw and
Pasti-Sorokin-Tonin actions are self-dual with respect to a common
anti-dualization of 1-form `field strengths' given by eq.(12) in $D=2$
dimensional space-time, and that they are self-dual with respect to another
common anti-dualization of 3-form field strengths given by eq.(24) in $D=6$
dimensional space-time. For the Srivastava action, we have verified that it
has self-duality under a generalized dual transform of 1-form `field strength'
expressed by eq.(42) in the $D=2$ case, and that it has self-duality under
another generalized dual transform of 3-form field strength expressed by
eq.(48) in the $D=6$ case. Here the word `generalized' means that
$G^{\mu}(\psi)$ and $F^{\mu}(\phi)$
do not relate with an anti-duality
$G^{\mu}(\psi)=-{\epsilon}^{{\mu}{\nu}}F_{\nu}(\phi)$
on the mass shell ${\cal F}^{\mu}(\phi)=0$ in $D=2$ dimensions, and that
$G^{{\mu}{\nu}{\sigma}}(B)$ and $F^{{\mu}{\nu}{\sigma}}(A)$
do not relate with another anti-duality
$G^{{\mu}{\nu}{\sigma}}(B)=-\frac{1}{3!}{\epsilon}^{{\mu}{\nu}{\sigma}{\rho}
{\eta}{\delta}}F_{{\rho}{\eta}{\delta}}(A)$
on the mass shell ${\cal F}^{{\mu}{\nu}{\sigma}}(A)=0$ in $D=6$ dimensions. The
reason is the linearity of the self-duality condition introduced with an
auxiliary field in the Srivastava action. We emphasize that this type of
anti-duality is not necessary for self-duality of actions because the
self-duality condition, i.e., the mass shell condition, cannot directly be
imposed on actions. Moreover, we have found a generally satisfied relation for
all the four actions discussed in this paper, that is, eq.(9) for the $D=2$ case
and eq.(20) for the $D=6$ case. This relation means that the self-duality
condition remains unchanged although the transforms of field strengths are
quite different from one action to another. Incidentally, we do not mention in
our paper the duality property of actions under transforms of auxiliary fields
because on one hand it is a trivial problem for the first three chiral {\em p}-form
actions, and on the other hand it has been studied in detail for the
Pasti-Sorokin-Tonin action [10]. The triviality is caused by the linearity of
auxiliary fields in the Siegel and Srivastava actions [7,8] and by the
non-existence of auxiliary fields in the Floreanini-Jackiw action [11,12].

    We have tried to extend the self-duality of actions from free theories to
interacting ones and chosen, as our example, the action of the
Floreanini-Jackiw chiral bosons interacting with gauge fields. By utilizing
the concept of the generalized dualization extracted from the self-duality
of the Srivastava action, we obtain that the action of the
interacting theory is self-dual with respect to a generalized anti-dualization
of the 1-form `field strength' of chiral scalars.

    As stated in Ref.[10] that the self-duality of the Pasti-Sorokin-Tonin
action remains in $D=2(p+1)$ dimensions, we can conclude that the Siegel,
Floreanini-Jackiw and Srivastava actions are also self-dual in $D=2(p+1)$
dimensional space-time. Finally, we point out that the self-duality also
exists in a wider context of theoretical models that relate to chiral {\em p}-forms, such as the generalized chiral Schwinger model (GCSM) [18], whose self-duality corresponds to the vector and axial vector current duality. 
This work is arranged in a separate paper [19].

{\em Note added.} The Kavalov-Mkrtchyan formulation [20] can be proved to be self-dual with respect to an anti-dualization of chiral 2-form fields along the line of this paper. We thank Dr. R. Manvelyan for pointing this out.
\vskip 20mm
\noindent
{\bf Acknowledgments}
\par
Y.-G. Miao is indebted to the Alexander von Humboldt Foundation for financial
support. He is also supported in part by the National Natural Science           Foundation
of China under grant No.19705007 and by the Ministry of Education of China
under the special project for scholars returned from abroad.

\end{document}